\documentclass[journal, draftcls, 12pt, a4paper, onecolumn]{IEEEtran}
\usepackage{ifpdf}
\usepackage{cite}
\ifCLASSINFOpdf
	\usepackage[pdftex]{graphicx}
	\DeclareGraphicsExtensions{.pdf,.jpeg,.png}
\else
	\usepackage[dvips]{graphicx}
	\DeclareGraphicsExtensions{.eps}
\fi
\usepackage[cmex10]{amsmath}
\usepackage{amssymb,amsfonts}
\usepackage{mathabx}

\usepackage{color}%
\usepackage{multicol}%
\usepackage{dcolumn,multirow}
\usepackage[normalem]{ulem}% \uline{underlined text}
\usepackage{comment}

\usepackage{algorithmic}
\usepackage{array}
\usepackage{mdwmath}
\usepackage{mdwtab}
\usepackage{eqparbox}
\usepackage[tight,footnotesize]{subfigure}
\usepackage{fixltx2e}
\usepackage{stfloats}
\usepackage{tikz}
\usetikzlibrary{shapes,arrows}

\ifCLASSOPTIONcaptionsoff
	\usepackage[nomarkers]{endfloat}
	\let\MYoriglatexcaption\caption
	\renewcommand{\caption}[2][\relax]{\MYoriglatexcaption[#2]{#2}}
\fi
\usepackage{url}
\newtheorem{theorem}{Theorem}
\newtheorem{lemma}[theorem]{Lemma}
\newtheorem{corollary}[theorem]{Corollary}
\newtheorem{definition}[theorem]{Definition}

\begin{document}
\title{New Decoding Scheme for LDPC Codes Based on Simple Product Code Structure}
\author{Beomkyu~Shin, Seokbeom~Hong, Hosung~Park, Jong-Seon~No,~\IEEEmembership{Fellow,~IEEE,} and~Dong-Joon~Shin,~\IEEEmembership{Senior~Member,~IEEE}% <-this % stops a space
\thanks{B.~Shin was with the Department of Electrical Engineering and Computer Science, INMC, Seoul National University, Seoul 151-744, Korea~(e-mail:thechi76@snu.ac.kr). He is now with the Samsung Electronics, Co., Ltd., Hwaseong, Gyeonggi-do, 445-701, Korea.}%
\thanks{S.~Hong, H.~Park, and J.-S.~No are with the Department of Electrical Engineering and Computer Science, INMC, Seoul National University, Seoul 151-744, Korea~(e-mail:fousbyus@ccl.snu.ac.kr, \{lovepk98, jsno\}@snu.ac.kr).}%
\thanks{D.-J. Shin is with the Department of Electronic Engineering, Hanyang University, Seoul 133-791, Korea~(e-mail: djshin@hanyang.ac.kr).}%
\ifCLASSOPTIONdraftcls
\else
\thanks{This work was supported by the Korea Science and Engineering Foundation (KOSEF) grant funded by the Korea government (MEST) (No. 2009-0081441).}%
\fi
%\thanks{Manuscript received February 20, 2010; revised April 12, 2010.}
}

\markboth{}%
{B. Shin \MakeLowercase{\textit{et al.}}: New Decoding Scheme for LDPC Codes Based on Simple Product Code Structure}


\maketitle

\begin{abstract}
In this paper, a new decoding scheme for low-density parity-check (LDPC) codes using the concept of simple product code structure is proposed based on combining two independently received soft-decision data for the same codeword. %The proposed scheme could be applied to any linear codes with ability for decoding-error detection with high probability such as LDPC codes. 
LDPC codes act as horizontal codes of the product codes and simple algebraic codes are used as vertical codes to help decoding of the LDPC codes. 
The decoding capability of the proposed decoding scheme is defined and analyzed using the parity-check matrices of vertical codes and especially the {\it combined-decodability} is derived for the case of single parity-check (SPC) and Hamming codes being used as vertical codes. 
It is also shown that the proposed decoding scheme achieves much better error-correcting capability in high signal to noise ratio (SNR) region with low additional decoding complexity, compared with a conventional decoding scheme.
\end{abstract}


% \fi
\IEEEpeerreviewmaketitle

\section{Introduction}
\IEEEPARstart{R}{ecently}, extremely low error probability has been required in many application areas such as wireless communication systems without feedback and data storage systems. Low-density parity-check (LDPC) codes\cite{Gallager},\cite{MacKay} have become one of promising error-correcting codes in these areas due to their near capacity-approaching performance. %Since low decoding complexity can be achieved by various iterative decoding algorithms, LDPC codes have been adopted in many practical applications.
However, since finite-length LDPC codes show a serious error-floor problem\cite{Richardson} in high signal to noise ratio (SNR) region, it may be difficult to achieve an extremely good error correction performance.

Many researches on the iterative decoding of LDPC codes (e.g. \cite{Fossorier} -\cite{Planjery2}) have been made to achieve good error correction performance by improving only the decoding algorithm while not changing the code structure. %However, there might be several applications with requirements for performance and complexity which is also not easy to be achieved by these schemes.
However, the requirement for the performance and complexity of decoder is getting more strict.
Thus, more powerful decoding schemes for LDPC codes are required.

%As indicated in \cite{Elias}, the product codes can be decoded by sequentially hard-decision decoding the rows and columns of codeword matrix in order to reduce the decoding complexity. However, for the optimal decoding performance, the maximum likelihood (ML) soft-decision decoding for the horizontal and vertical codes must be performed. Thus, soft-input/soft-output decoders are needed to decode the rows and columns of codeword matrix in $\mathcal{P}$. However, the ML decoding of the product codes is too complicated for practical use. Moreover, it requires two decoders if the vertical code is different from the horizontal code, which increases the hardware complexity.

In this paper, we propose a decoding scheme for LDPC codes using simple product code structure to improve the error correction performance together with error floor. The proposed decoding scheme is based on combining two independently received soft-decision data for an unsuccessfully decoded codeword, which achieves good decoding performance with relatively low additional decoding complexity.
In the encoding procedure of the proposed scheme, codewords of the LDPC codes are stacked in rows as horizontal codewords of the product codes. Then vertical codes which are used to help decoding of the LDPC codewords are applied to the stack of LDPC codewords, which results in a codeword matrix like product code structure.
In the decoding procedure, decoding of each horizontal LDPC codeword in a codeword matrix is performed firstly according to decoding algorithm of LDPC codes.
If some codewords fail to be successfully decoded, re-decoding based on combining two received soft-decision data for the same codeword will be performed.

Also, we define and analyze the decoding capability of the proposed decoding scheme using the parity-check matrices of vertical codes and derive the {\it combined-decodability} for the case of single parity-check (SPC) and Hamming codes being used as vertical codes. Especially, the proposed decoding scheme shows better error-correcting capability in high SNR region. 

Any linear codes can be used as vertical codes in the proposed scheme. However, since vertical codes are only used to help decoding of LDPC codes in the proposed scheme, short codes with high rate are preferable as vertical codes. Thus, simple algebraic codes such as SPC or Hamming codes are appropriate for the vertical codes.
Even though the vertical codes are assumed to be systematic in this paper, they do not have to be systematic for the proposed decoding scheme.

\begin{figure}[ht!]
	\centering
	\includegraphics[width=\columnwidth]{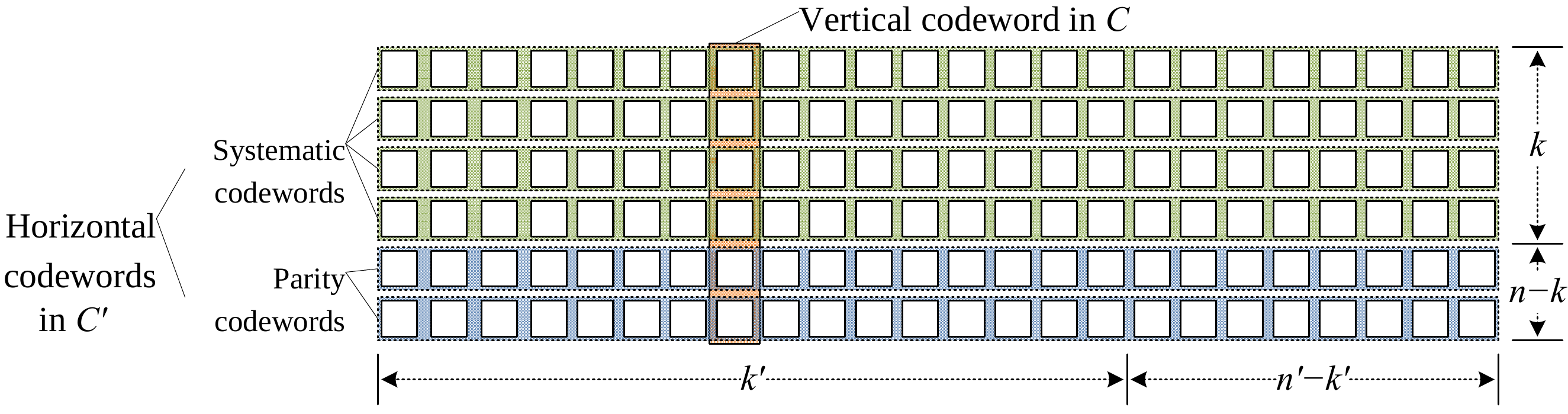}
	\caption{Structure of a codeword matrix of product code.}
    \label{fig:StructureProd}
    \vspace{-8mm}
\end{figure}

The proposed decoding scheme can be applied to any other linear codes with soft-decision decoding than LDPC codes. However, undetected errors in the horizontal codewords cause a serious problem for the proposed scheme. If undetected errors seldom occur and soft-decision decoding is used for the horizontal codes such as LDPC and turbo codes\cite{Berrou}, the proposed decoding scheme can be effectively applied.

This paper is organized as follows: In Section~\ref{sec:ProposedScheme}, after introducing and defining some notations and operations, a new decoding scheme for LDPC codes using simple product code structure is proposed. Two simple examples for the proposed decoding scheme are given in Section~\ref{sec:SimpleExamples}. In Section~\ref{sec:CorrectionCapability}, the combined-decodability of the proposed decoding scheme is analyzed. The numerical results are shown in Section~\ref{sec:SimulationResults} and the conclusion is given in Section~\ref{sec:Conclusions}.

\section{A New Decoding Scheme Based on Simple Product Code Structure}\label{sec:ProposedScheme}

In this section, we propose a new decoding scheme for LDPC codes using the concept of product codes. In the proposed scheme, LDPC codes act as horizontal codes of the product codes and simple codes such as SPC or Hamming codes are used for vertical codes of the product code to help the decoding of the horizontal codes.

\subsection{Notations and Basic Operations}

Product codes are serially concatenated codes \cite{MacWilliams}, which were introduced in \cite{Elias}. By using the concept of product codes, a long block code can be easily constructed by using two or more short block codes. Now, we consider two systematic linear block codes, $\mathcal{C}$ with parameters $(n, k, d_{\rm min} )$ and $\mathcal{C}'$ with parameters $(n', k', d'_{\rm min})$, where $n (\textrm{or~} n'), k (\textrm{or~} k')$, and $d_{\rm min} (\textrm{or~} d'_{\rm min})$ denote the code length, the number of information bits, and the minimum Hamming distance, respectively. The product code $\mathcal{P} = \mathcal{C} \otimes \mathcal{C} '$ is obtained by the following steps as illustrated in Fig.~\ref{fig:StructureProd}:
\begin{enumerate}
	\item[1)] Placing $k \times k'$ information bits in an array of $k$ rows and $k'$ columns;
	\item[2)] Encoding each of $k$ rows using code $\mathcal{C}'$;
	\item[3)] Encoding each of $n'$ columns using code $\mathcal{C}$.
\end{enumerate}
The codes $\mathcal{C}$ and $\mathcal{C}'$ are also called as vertical code and horizontal code of the product code $\mathcal{P}$, respectively. The parameters of the product code $\mathcal{P}$ are $(n \times n', k \times k', d_{\rm min} \times d'_{\rm min})$ and the code rate is given as $R \times R'$, where $R$ and $R'$ are the code rates of $\mathcal{C}$ and $\mathcal{C}'$, respectively.

In this paper, for simplicity, only binary block codes are considered but the proposed decoding scheme can be directly applied to the nonbinary codes.
Let $M = 2^m-1$ and $e$, $0 \leq e \leq n$, be the number of unsuccessfully decoded horizontal codewords in a codeword matrix after first decoding of each $n$ horizontal codewords.
Parity-check matrix and its modifications of a vertical code in the product code are defined as:
\begin{itemize}
	\item $H$: $m \times n$ parity-check matrix of a vertical code in the product code; 
	\item $H_E$: $M \times n$ extended parity-check matrix whose rows are all possible nonzero linear combinations of rows in $H$;
	\item $H_P$: $M \times e$ punctured parity-check matrix constructed by selecting $e$ columns from $H_E$.
\end{itemize}
%After decoding each of $n$ horizontal codewords, some codewords may fail to be successfully decoded. Let $e$, $0 \leq e \leq n$, be the number of unsuccessfully decoded horizontal codewords in a codeword matrix.
It can be assumed that $H$ contains no all-zero column. It is clear that the length $n$ of the vertical code also denotes the number of horizontal codewords in a codeword matrix of the product code.
The column indices of $H_P$ are same as the row indices of the unsuccessfully decoded horizontal codewords in the codeword matrix.

Let $\mathbf{c}_i = (c_{i1},c_{i2}, \ldots, c_{in'} )$, $1 \leq i \leq n$, denote $n$ binary horizontal codewords and $\mathbf{r}_i = (r_{i1},r_{i2}, \ldots, r_{in'} )$ denote the {\it soft-decision vector} for $\mathbf{c}_i$, where $r_{ij}$ is the log likelihood ratio (LLR) for $c_{ij}$ calculated from the corresponding received data $a_{ij}$ defined by 
\begin{equation*}
	r_{ij} = \log \left (\frac{\mathrm{Pr}(a_{ij}|c_{ij}=+1)}{\mathrm{Pr}(a_{ij}|c_{ij}=-1)} \right).
\end{equation*}
Then, the check equations (or rows) in $H_E=[h_{ji}]$ describe all possible relations among $\mathbf{c}_i$'s, that is, $\sum_{i=1}^{n} h_{ji} \cdot \mathbf{c}_i=\mathbf{0}, 1\leq j \leq M$. Note that these check equations (or rows) of $H_E$ are not linearly independent. Since the vertical code is a systematic code, we can use $\mathbf{s}_i = (s_{i1},s_{i2}, \ldots, s_{in'} )$, $1 \leq i \leq k$, and $\mathbf{p}_i = (p_{i1},p_{i2}, \ldots, p_{in'} ), 1 \leq i \leq n-k$, to denote the {\it systematic (horizontal) codewords} including the systematic part of the vertical code and the {\it parity (horizontal) codewords} including the parity part of the vertical code as given in Fig.~\ref{fig:StructureProd}, respectively.

Special algebra for LLR values in \cite{Hagenauer} can be used with small modification. We use the operation $\boxplus$ for combining two LLR values defined as
\begin{equation}\label{eq:boxplus}
	r_{ij} \boxplus r_{i'j} = 2 \textrm{tanh}^{-1} \left ( \textrm{tanh} \left ( \frac{r_{ij}}{2} \right ) \textrm{tanh} \left (\frac{r_{i'j}}{2} \right ) \right ).
\end{equation}
Let $ \mathbf{\bar r}_i = (\bar r_{i1}, \bar r_{i2}, \ldots, \bar r_{in'} )$, $1 \leq i \leq n$, denote the vector of LLR values computed from the horizontal codeword $\mathbf{c}_i$ itself, i.e., $\bar r_{ij} = - \infty$ if $c_{ij}=1$ and $\bar r_{ij} = +\infty$ if $c_{ij}=0$. Then it is easy to show that $r_{ij} \boxplus \bar{r}_{i'j}$ can be calculated using the sign change operation as
\begin{equation}\label{eq:extendedboxplus}
	\begin{cases}
		r_{ij},		&\textrm{~if~}\bar r_{i'j} = +\infty\\
		-r_{ij},	&\textrm{~if~} \bar r_{i'j} = - \infty.
	\end{cases}
\end{equation}

\subsection{A New Decoding Scheme}
%In this subsection, we propose a new decoding scheme for the product codes, where the horizontal codes have the properties of linearity and decoding-error detection capability.

In the proposed decoding scheme, the received codewords of the LDPC codes are stacked as horizontal codewords for the product code. The number of codewords in a stack is determined by the vertical codes which are used to help decoding of the horizontal codes. Next, we attempt to decode each horizontal codeword in the codeword matrix. If all the systematic codewords $\mathbf{s}_i$, $1 \le i \le k$, are successfully decoded, the decoding process will be successfully finished. However, if some horizontal codewords fail to be decoded, then re-decoding of the horizontal codewords using $H_P$ of the vertical codes will be performed. The details of this procedure are explained below.

Suppose that there are $e$ unsuccessfully decoded horizontal codewords in a codeword matrix and the $j$-th row of $H_P$ of the vertical codes has the minimum nonzero Hamming weight $e_{\rm min}$ among all rows of $H_P$. Clearly, it implies that $e_{\rm min}$ unsuccessfully decoded horizontal codewords corresponding to the nonzero elements are involved in the $j$-th check equation of $H_E$ of the vertical code, which is the check equation affected by the minimum number of unsuccessfully decoded horizontal codewords. Therefore, $H_P$ is used to decide which check equation of $H_E$ is most easily solvable, i.e., which check equation contains the fewest unsuccessfully decoded horizontal codewords. In the proposed decoding scheme, the check equation which contains the fewest unsuccessfully decoded horizontal codewords is first utilized for re-decoding.
 
According to the value of $e_{\rm min}$, the proposed decoding scheme is performed by considering the following three cases.

Case 1) $e_{\rm min} = 1$:

If there is only one unsuccessfully decoded horizontal codeword participating in a check equation of $H_E$, say the $j$-th row of $H_E$, of the vertical code, we can correct it simply by the modulo 2 addition (denoted by $\oplus$) of the other successfully decoded horizontal codewords participating in the same check equation. Let $\mathbf{c}_i$ be the unsuccessfully decoded horizontal codeword participating in the $j$-th check equation of $H_E$ of the vertical code. Then we can recover $\mathbf{c}_i$ by 
\begin{equation}\label{eq:oneerror}
	\mathbf{c}_i = \bigoplus_{\forall l \in H_E(j) \backslash \{i\}} \mathbf{c}_l
\end{equation}
where $H_E(j)$ is the set of indices of the nonzero elements in the $j$-th row of $H_E$. As $\mathbf{c}_i$ is successfully decoded, the corresponding column of $H_P$ is removed and the number of columns in $H_P$ reduces by one. This process is repeatedly applied to the check equations which satisfy the $e_{\rm min} =1$ condition for updated $H_P$. 

Case 2) $e_{\rm min} = 2$:

If there are only two unsuccessfully decoded horizontal codewords participating in a check equation (i.e., a row) of $H_E$ of the vertical code, we can re-decode the combined vector of two independently received soft-decision vectors for the same horizontal codeword similar to the type-I hybrid-automatic repeat request (H-ARQ) scheme \cite{Lin}. We already know that among horizontal codewords participating in a check equation of $H_E$, any codeword is equivalent to the sum of all the other codewords. Thus we can derive two independently received soft-decision vectors for any of two unsuccessfully decoded horizontal codewords; (a) one from an unsuccessfully decoded horizontal codeword, (b) the other one from the soft-decision vector of the other unsuccessfully decoded horizontal codeword by changing signs of its elements according to the successfully decoded horizontal codewords participating in the same check equation.

Let $\mathbf{c}_{i_1}$ and $\mathbf{c}_{i_2}$ be the unsuccessfully decoded horizontal codewords participating in the $j$-th check equation of $H_E$ of the vertical code. Then, from~\eqref{eq:oneerror}, we can have
\begin{equation}\label{eq:twoerrors}
	\mathbf{c}_{i_1} = \mathbf{c}_{i_2} \oplus \left( \bigoplus_{\forall l \in H_E(j) \backslash \{i_1, i_2\}} \mathbf{c}_l \right).
\end{equation}

In addition to the soft-decision vector $\mathbf{r}_{i_1}$ for $\mathbf{c}_{i_1}$, we can derive the other soft-decision vector for $\mathbf{c}_{i_1}$ by using the relation \eqref{eq:twoerrors} as
\begin{equation}\label{eq:LLRtwoerrors}
	\mathbf{r}_{i_2} \boxplus \left ( \bigboxplus_{\forall l \in H_E(j) \backslash \{i_1, i_2\}} \mathbf{\bar r}_l \right ).
\end{equation}
Since it can be regarded that the horizontal codeword $\mathbf{c}_{i_1}$ is independently received twice through the channel, we can just add two soft-decision vectors as
\begin{equation}\label{eq:LLRsum}
	\mathbf{r}_{i_1} + \left [ \mathbf{r}_{i_2} \boxplus \left( \bigboxplus_{\forall l \in H_E(j) \backslash \{i_1, i_2\}} \mathbf{\bar r}_l \right) \right]
\end{equation}
and re-decode it by achieving approximately 3dB performance gain similar to type-I H-ARQ scheme.
% It is well known that re-decoding with the combined soft-decision vector from two independently received soft-decision vectors achieves about 3dB performance gain. In this case, perfect independentness between two soft-decision vectors from unsuccessfully decoded horizontal codewords is not guaranteed, but we confirm by numerical analysis that performance gain is almost 3dB with combining soft-decision vectors. 
If the re-decoding is successful, then the $j$-th check equation of $H_E$ contains only one unsuccessfully decoded horizontal codeword $\mathbf{c}_{i_2}$ and it can be recovered by the method in Case 1).

Case 3) $e_{\rm min} \geq 3$:

If there are three or more unsuccessfully decoded horizontal codewords participating in a check equation of $H_E$ of the vertical code, combining the soft-decision vectors for further re-decoding is still possible. First of all, the unsuccessfully decoded horizontal codewords participating in the same check equation of $H_E$ are partitioned into two groups and a soft-decision vector for each group is obtained by summing all the unsuccessfully decoded soft-decision vectors in that group using the operation \eqref{eq:boxplus}. Then, for these two soft-decision vectors, the same method for two unsuccessfully decoded horizontal codewords in Case~2) is applied. We can try this process to all possible partitioning of unsuccessfully decoded horizontal codewords until the decoding succeeds and then reduces the number of unsuccessfully decoded codewords in the same check equation. However, it should be noted that more combined horizontal codewords result in less performance gain.

Let $\mathbf{c}_{i_1}, \cdots, \mathbf{c}_{i_{g}}, \cdots, \mathbf{c}_{i_{h}},\cdots, \mathbf{c}_{i_{e_\textrm{min}}}$ be the $e_\textrm{min}$ unsuccessfully decoded horizontal codewords participating in the $j$-th check equation of $H_E$ of the vertical code. Then \eqref{eq:oneerror} can be rewritten as
\begin{equation}\label{eq:moreerrors}
	\left ( \bigoplus_{\forall l \in \{i_{h+1}, \cdots, i_{e_\textrm{min}}\}} \mathbf{c}_l \right ) = \left ( \bigoplus_{\forall l \in \{i_1, \cdots, i_{h} \}} \mathbf{c}_l \right ) \oplus \left ( \bigoplus_{\forall l \in H_E(j) \backslash \{i_1, \cdots, i_{e_\textrm{min}}\}} \mathbf{c}_l \right ).
\end{equation}
Two soft-decision vectors for $\bigoplus_{\forall l \in \{i_{h+1}, \cdots, i_{e_\textrm{min}} \}} \mathbf{c}_l$ can be obtained as
\begin{equation}
		\bigboxplus_{\forall l \in \{i_{h+1}, \cdots, i_{e_\textrm{min}}\}} \mathbf{r}_l  \label{eq:LLRgroup1}
\end{equation}
and
\begin{equation}
		\left ( \bigboxplus_{\forall l \in \{i_1, \cdots, i_{h}\}} \mathbf{r}_l \right )  \boxplus \left ( \bigboxplus_{\forall l \in H_E(j) \backslash \{i_1, \cdots, i_{e_\textrm{min}}\}} \mathbf{\bar r}_l \right ). \label{eq:LLRgroup2}
\end{equation}
Since it can be regarded that one horizontal codeword is independently received twice through the channel, we can add two soft-decision vectors in \eqref{eq:LLRgroup1} and \eqref{eq:LLRgroup2} and re-decode it. If the re-decoding succeeds, then $\bigoplus_{\forall l \in \{i_1, \cdots, i_{h} \}} \mathbf{c}_l$ and $\bigoplus_{\forall l \in \{i_{h+1}, \cdots, i_{e_\textrm{min}}\}} \mathbf{c}_l$ are known. By repeating the previous process, $\mathbf{c}_{i_1}, \cdots, \mathbf{c}_{i_{g}}, \cdots, \mathbf{c}_{i_{h}}$ can also be split into two groups as
\begin{equation}\label{eq:reducederrors}
	\left ( \bigoplus_{\forall l \in \{i_{g+1}, \cdots, i_{h} \}} \mathbf{c}_l \right ) = \left ( \bigoplus_{\forall l \in \{i_{1}, \cdots, i_{g}\}} \mathbf{c}_l \right ) \oplus \left ( \bigoplus_{\forall l \in H_E(j) \backslash \{i_1, \cdots, i_{h}\}} \mathbf{c}_l \right )
\end{equation}
and the same process can be applied to decode $\bigoplus_{\forall l \in \{i_{g+1}, \cdots, i_{h} \}} \mathbf{c}_l$. If the re-decoding is unsuccessful, then different grouping of the unsuccessfully decoded codewords can be tried. This process is repeatedly performed until all errors are corrected.

\begin{figure}[t!]
	\centering
	\includegraphics[width=\columnwidth]{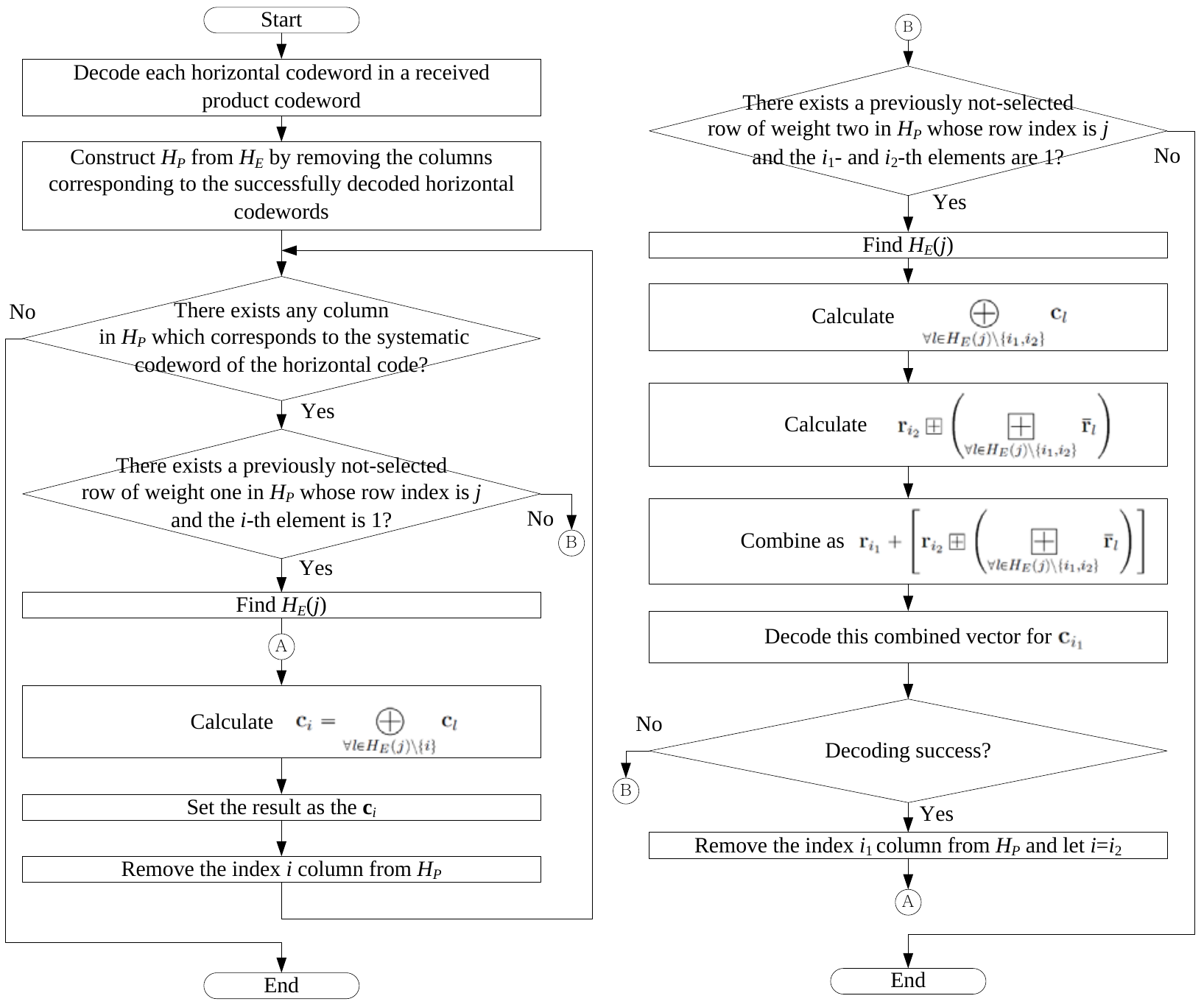}
	\caption{Flowchart of the proposed decoding process.}
	\label{fig:flowchart_main}
	\vspace{-8mm}
\end{figure}

The flowchart of the proposed decoding scheme is shown in Fig.~\ref{fig:flowchart_main}, where only $e_{\rm min} \leq 2$ is considered. This flowchart is based on the assumption that the vertical code is a linear systematic code. Note that four steps in the left bottom part of Fig.~\ref{fig:flowchart_main} are equivalent to the erasure decoding under binary erasure channel. It is well known that the {\em error correctability} $\tau$ of the vertical code is determined by the minimum Hamming distance $d_{\rm min}$ of the code, i.e., $\tau = d_{\rm min} - 1$ under binary erasure channel.

%The improvement of error correction capability of the proposed decoding scheme can be estimated by the following approximation. Let $p$ denote the codeword error rate of the horizontal code, i.e., punctured LDPC code. Then the probability ${\rm Pr}\{e\}$ of $e$ horizontal codeword errors in a codeword matrix is given as
%\begin{equation*}
%		{\rm Pr}\{ e\} = \binom{n}{e}p^e (1-p)^{n-e}.
%\end{equation*}
%Let ${\rm WER}(e)$ denote the contribution of $e$ horizontal codeword errors in a codeword matrix to the word error rate (${\rm WER}$) $p$ of each horizontal codeword, that is, 
%\begin{equation*}
%	{\rm WER}(e) = \frac{e{\rm Pr}\{e\}}{n} = \frac{e}{n} \binom{n}{e}p^e (1-p)^{n-e}.
%\end{equation*}
%Then the validity of ${\rm WER}(e)$ can be easily checked by
%\begin{equation*}
%		\sum_{e=1}^{n} {\rm WER}(e) = \sum_{e=1}^{n} \frac{e{\rm Pr}\{e\}}{n} = \frac{np}{n} = p.
%\end{equation*}
%Let ${\rm WER}_i$ denote the ${\rm WER}$ when the perfect error correction is assumed for all error patterns of $i$ or less horizontal codeword errors. Assume that the horizontal codeword errors are occurred independently. Then, 
%\begin{equation*}
%	{\rm WER}_i	\triangleq p - \sum_{e=1}^{i} {\rm WER}(e).
%\end{equation*}
%Especially for $i = 2$, we can have the approximation as
%\begin{equation}\label{eq:approximation}
%	{\rm WER}_2 \simeq (n-1)(n-2)p^3.
%\end{equation}
%From~\eqref{eq:approximation}, more performance improvement can be expected in lower error rate region even in the case of $i = 2$.

\section{Examples}\label{sec:SimpleExamples}
\subsection{SPC Codes as Vertical Codes}
Assume that an SPC code with length $n$ is used as a vertical code of the product code and an LDPC code is used as a horizontal code. Then, the parity-check matrix $H$ of the vertical code and its $H_E$ can be written as
\begin{equation*}
	H = H_E =
	\begin{bmatrix}
		1 & 1 & \cdots & 1 & 1\\
	\end{bmatrix}.
\end{equation*}
From the parity-check matrix, it holds
\begin{equation*}
	\left( \bigoplus_{1 \leq i \leq n-1} \mathbf{s}_i \right) \oplus \mathbf{p}_1 = \mathbf{0}.
\end{equation*}
After decoding each of $n$ horizontal codewords, there may exist $e$ unsuccessfully decoded horizontal codewords. Since $H$, $H_E$ , and $H_P$ have only one row whose entries are all 1, $e_{\rm min}$ is always equal to $e$. If $e_{\rm min}=1$ and the unsuccessfully decoded codeword is parity codeword $\mathbf{p}_1$, we already successfully decode the whole systematic horizontal codewords $\mathbf{s}_1 \sim \mathbf{s}_{n-1}$. If $e_{\rm min}=1$ and the unsuccessfully decoded codeword is a systematic codeword $\mathbf{s}_i$, this codeword can be easily corrected by the even parity condition for the vertical code as
\begin{equation}\label{eq:6_oneerror}
	\mathbf{s}_i = \left( \bigoplus_{1\leq l \leq n-1, l\neq i} \mathbf{s}_l \right) \oplus \mathbf{p}_1.
\end{equation}
If $e_{\rm min}=2$ and two systematic codewords $\mathbf{s}_{i_1}$ and $\mathbf{s}_{i_2}$ are unsuccessfully decoded, \eqref{eq:6_oneerror} can be rewritten as
\begin{equation*}
	\mathbf{s}_{i_1} = \left( \bigoplus_{1\leq l \leq n-1, l\neq i_1} \mathbf{s}_l \right) \oplus \mathbf{p}_1 = \mathbf{s}_{i_2} \oplus \left( \bigoplus_{1\leq l \leq n-1, l\neq i_1, i_2} \mathbf{s}_l \right) \oplus \mathbf{p}_1. 
\end{equation*}
We can exactly calculate $\left( \bigoplus_{\hspace{2mm} 1\leq l \leq n-1, l\neq i_1, i_2} \mathbf{s}_i \right) \oplus \mathbf{p}_1$ because all these codewords are successfully decoded. With this result and the soft-decision vector for $\mathbf{s}_{i_2}$, we have another independently received soft-decision vector $\mathbf{r'}_{i_1}$ for a codeword $\mathbf{s}_{i_1}$ as 
\begin{equation*}
	\mathbf{r'}_{i_1} = \mathbf{r}_{i_2} \boxplus \left( \bigboxplus_{1\leq l \leq n, l\neq i_1, i_2} \mathbf{\bar r}_l \right).
\end{equation*}
Similar to the type-I H-ARQ, we can add two soft-decision vectors $\mathbf{r}_{i_1}$ and $\mathbf{r'}_{i_1}$ and then re-decode it to obtain $\mathbf{s}_{i_1}$. If one of two unsuccessfully decoded codewords is $\mathbf{p}_1$, it is easy to show that the same procedure can be applied.

If $e_{\rm min}=3$ and three systematic codewords $\mathbf{s}_{i_1}$, $\mathbf{s}_{i_2}$, and $\mathbf{s}_{i_3}$ are unsuccessfully decoded, \eqref{eq:6_oneerror} can be rewritten as
\begin{equation*}
	\mathbf{s}_{i_1}	= \left( \bigoplus_{1\leq l \leq n-1, l\neq i_1} \mathbf{s}_l \right) \oplus \mathbf{p}_1 
				= \mathbf{s}_{i_2} \oplus \mathbf{s}_{i_3} \oplus \left( \bigoplus_{1\leq l \leq n-1, l\neq i_1, i_2, i_3} \mathbf{s}_l \right) \oplus \mathbf{p}_1. 
\end{equation*}
Then, we can combine the soft-decision vectors for $\mathbf{s}_{i_2}$ and $\mathbf{s}_{i_3}$ by \eqref{eq:boxplus} and exactly calculate $\left( \bigoplus_{\hspace{2mm} 1\leq l \leq n-1, l\neq i_1, i_2, i_3} \mathbf{s}_l \right) \oplus \mathbf{p}_1$. By using these vectors, we derive another independently received soft-decision vector $\mathbf{r'}_{i_1}$ for the codeword $\mathbf{s}_{i_1}$ as
\begin{equation*}
	\mathbf{r'}_{i_1} = \mathbf{r}_{i_2} \boxplus \mathbf{r}_{i_3} \boxplus \left( \bigboxplus_{1\leq l \leq n, l\neq i_1, i_2, i_3} \mathbf{\bar r}_l \right).
\end{equation*}
Finally, we add two soft-decision vectors $\mathbf{r}_{i_1}$ and $\mathbf{r'}_{i_1}$ and re-decode it for $\mathbf{s}_{i_1}$. If the re-decoding fails, then we can retry the same decoding procedure for $\mathbf{s}_{i_2}$ or $\mathbf{s}_{i_3}$ instead of $\mathbf{s}_{i_1}$. In the case of $e_{\rm min} \geq 4$, the similar process can be applied.

\subsection{$(7,4)$ Hamming Codes as Vertical Codes}\label{sec:ExampleHamming}
Assume that a $(7,4)$ Hamming code is used as a vertical code of the product code and an LDPC code is used as a horizontal code. Then, the parity-check matrix $H$ of the vertical code and its $H_E$ can be written as
\begin{equation*}
	H = 
	\begin{bmatrix}
		1 & 0 & 1 & 1 & 1 & 0 & 0 \\[-3mm]
		1 & 1 & 1 & 0 & 0 & 1 & 0 \\[-3mm]
		0 & 1 & 1 & 1 & 0 & 0 & 1 \\
	\end{bmatrix} \Rightarrow
	H_E = 
	\begin{bmatrix}
		1 & 0 & 1 & 1 & 1 & 0 & 0 \\[-3mm]
		1 & 1 & 1 & 0 & 0 & 1 & 0 \\[-3mm]
		0 & 1 & 1 & 1 & 0 & 0 & 1 \\[-3mm]
		0 & 1 & 0 & 1 & 1 & 1 & 0 \\[-3mm]
		1 & 1 & 0 & 0 & 1 & 0 & 1 \\[-3mm]
		1 & 0 & 0 & 1 & 0 & 1 & 1 \\[-3mm]
		0 & 0 & 1 & 0 & 1 & 1 & 1 \\
	\end{bmatrix}.
\end{equation*}

Then, the rows of $H_E$ give the following check equations for 7 horizontal codewords $\mathbf{s}_1$, $\mathbf{s}_2$, $\mathbf{s}_3$, $\mathbf{s}_4$, $\mathbf{p}_1$, $\mathbf{p}_2$, and $\mathbf{p}_3$;

\begin{align}
	\mathbf{s}_1 \oplus \mathbf{s}_3 \oplus \mathbf{s}_4 \oplus \mathbf{p}_1 &= \mathbf{0} \label{eq:hamming1} \\
	\mathbf{s}_1 \oplus \mathbf{s}_2 \oplus \mathbf{s}_3 \oplus \mathbf{p}_2 &= \mathbf{0} \label{eq:hamming2} \\ 
	\mathbf{s}_2 \oplus \mathbf{s}_3 \oplus \mathbf{s}_4 \oplus \mathbf{p}_3 &= 
\mathbf{0} \label{eq:hamming3} \\
	\mathbf{s}_2 \oplus \mathbf{s}_4 \oplus \mathbf{p}_1 \oplus \mathbf{p}_2 &= \mathbf{0} \label{eq:hamming4} \\
	\mathbf{s}_1 \oplus \mathbf{s}_2 \oplus \mathbf{p}_1 \oplus \mathbf{p}_3 &= \mathbf{0} \label{eq:hamming5}\\ 
	\mathbf{s}_1 \oplus \mathbf{s}_4 \oplus \mathbf{p}_2 \oplus \mathbf{p}_3 &= \mathbf{0} \label{eq:hamming6}\\
	\mathbf{s}_3 \oplus \mathbf{p}_1 \oplus \mathbf{p}_2 \oplus \mathbf{p}_3 &= \mathbf{0}. \label{eq:hamming7}
\end{align}

After decoding each of seven horizontal codewords, there may exist $e$ horizontal codewords, $0 \leq e \leq 7$, which fail to be correctly decoded. If $e=1$, at least one check equation from \eqref{eq:hamming1}$\sim$\eqref{eq:hamming7} contains that unsuccessfully decoded horizontal codeword, which can be easily corrected. It is easy to see that selecting proper check equation to correct an erroneous horizontal codeword is equivalent to finding a weight-1 row in $H_P$.

The same approach can be used to correct more than one error. For example, let $\mathbf{s}_1$, $\mathbf{s}_2$, and $\mathbf{s}_3$ be three unsuccessfully decoded horizontal codewords. Then, we have
\begin{equation*}
H_P=	\begin{bmatrix}
		1 & 0 & 1 \\[-3mm]
		1 & 1 & 1 \\[-3mm]
		0 & 1 & 1 \\[-3mm]
		0 & 1 & 0 \\[-3mm]
		1 & 1 & 0 \\[-3mm]
		1 & 0 & 0 \\[-3mm]
		0 & 0 & 1 \\
	\end{bmatrix}
\end{equation*}
and each recovery process for $\mathbf{s}_1$, $\mathbf{s}_2$, and $\mathbf{s}_3$ can be conducted using \eqref{eq:hamming6}, \eqref{eq:hamming4}, and \eqref{eq:hamming7}, which correspond to the sixth, the fourth, and the last rows of $H_P$, respectively. 

However, for three unsuccessfully decoded codewords $\mathbf{s}_1$, $\mathbf{s}_2$, and $\mathbf{s}_4$, there is no weight-1 row in the following $H_P$ constructed by selecting the first, second, and fourth columns of $H_E$, which has $e_{\rm min} = 2$ as
\begin{equation*}
H_P=	\begin{bmatrix}
		1 & 0 & 1 \\[-3mm]
		1 & 1 & 0 \\[-3mm]
		0 & 1 & 1 \\[-3mm]
		0 & 1 & 1 \\[-3mm]
		1 & 1 & 0 \\[-3mm]
		1 & 0 & 1 \\[-3mm]
		0 & 0 & 0 \\
	\end{bmatrix}.
\end{equation*}
Instead, as the Case 2) in the subsection \ref{sec:ProposedScheme}-B, by using \eqref{eq:hamming1} corresponding to the first row of the above $H_P$, we can modify the sign of each element of the soft-decision vector of $\mathbf{s}_4$ using the correctly decoded codewords $\mathbf{s}_3$ and $\mathbf{p}_1$, which results in another soft-decision vector for $\mathbf{s}_1$. This newly created soft-decision vector can be added to the original soft-decision vector for $\mathbf{s}_1$ and the re-decoding can be performed for this combined vector. If the re-decoding fails, we can try the same decoding procedure for another check equation corresponding to a weight-2 row in $H_P$.

Now, we consider more severe case, for example, there is only one successfully decoded horizontal codeword in a codeword matrix ($e=6$), say $\mathbf{s}_3$. Using \eqref{eq:hamming1}, we have the following re-decoding steps:
\begin{enumerate}
	\item Combine the soft-decision vectors of $\mathbf{s}_4$ and $\mathbf{p}_1$ using the operation \eqref{eq:boxplus} to have the soft-decision vector for $\mathbf{s}_4 \oplus \mathbf{p}_1$;
	\item Modify the sign of each element of the resulting soft-decision vector using $\mathbf{s}_3$ to obtain the soft-decision vector for $\mathbf{s}_3 \oplus \mathbf{s}_4 \oplus \mathbf{p}_1$.
\end{enumerate}
The resulting soft-decision vector for $\mathbf{s}_3 \oplus \mathbf{s}_4 \oplus \mathbf{p}_1$ can be considered as the other independently received vector for $\mathbf{s}_1$. Thus this newly created soft-decision vector can be added to the original soft-decision vector for $\mathbf{s}_1$ and the re-decoding can be performed for this combined soft-decision vector. If the decoding is successful, then it gives $e=5$ case. If the decoding is not successful, we can also try other check equation until the decoding is successful or there is no remaining check equation available.

\section{Combined-Decodability of the Proposed Decoding Scheme}\label{sec:CorrectionCapability}
In this section, the combined-decodability of the proposed decoding scheme is defined and the relation between the combined-decodability of the proposed decoding scheme and the vertical code with parity-check matrix $H$ is investigated.
Although the soft information combining of the proposed decoding scheme can be applied for any value of $e_{\rm min}$, the analysis of combined-decodability is performed by only considering the cases of $e_{\rm min} \le 2$ in this section.

\subsection{Definitions and Properties of Combined-Decodability}
\begin{definition}[$\epsilon$ combinable]
For a given $H$ of a vertical code $\mathcal{C}$ and a fixed $\epsilon$, if every possible $M \times \epsilon$ matrix $H_P$ contains at least one row of Hamming weight one or two, the vertical code $\mathcal{C}$ is said to be $\epsilon$ {\it combinable}.
%a product code $\mathcal{P} = \mathcal{C} \otimes \mathcal{C} '$ is said to be $\epsilon$ {\it combinable}. For simplicity, the vertical code $\mathcal{C}$ is also said to be $\epsilon$ {\it combinable}.
\end{definition}
\vspace{6mm}

\begin{definition}[$\eta$ combined-decodable]
If a vertical code $\mathcal{C}$ is $\epsilon$ combinable for all $\epsilon \leq \eta$, then $\mathcal{C}$ is said to be $\eta$ {\it combined-decodable}. If $\mathcal{C}$ is not $(\eta +1)$ combined-decodable but $\eta$ combined-decodable, then $\eta$ is called the combined-decodability of $\mathcal{C}$.
\end{definition}
\vspace{6mm}
 
Using the above definitions, the length of vertical code can be bounded as in the following lemmas and theorems.\\
 
\begin{lemma}
If a vertical code $\mathcal{C}$ with a parity-check matrix $H$ is $\epsilon$ combinable, any vertical code $\tilde{\mathcal{C}}$ whose parity-check matrix $\tilde{H}$ is constructed by selecting $\epsilon$ or more columns from $H$ is also $\epsilon$ combinable.
\end{lemma}\vspace{6mm}

\begin{proof}
A set of all possible $\tilde{H}''$'s which are constructed by selecting $\epsilon$ columns from $\tilde{H}'$ is contained in the set of all possible $H_P$'s which are constructed by selecting $\epsilon$ columns from $H_E$.
\end{proof}
\vspace{6mm}

\begin{lemma}
All vertical codes $\mathcal{C}$ are $1$ and $2$ combinable.
\end{lemma}\vspace{6mm}
\begin{proof}
Since there is no all-zero column in the parity-check matrix $H$ of any vertical code $\mathcal{C}$, all vertical codes $\mathcal{C}$ are both $1$ and $2$ combinable.
\end{proof}\vspace{6mm}
\begin{lemma}
If a vertical code $\mathcal{C}$ with an $m \times n$ parity-check matrix $H$ is $3$ combinable, then $n \leq 2M$, where $M=2^m -1$. Moreover, when $n=2M$, $\mathcal{C}$ is $3$ combinable if and only if $H$ is constructed by using every nonzero $m$-tuple column vector exactly twice as its columns.
\end{lemma}\vspace{6mm}
\begin{proof}
If $n \geq 2M +1$, there should exist three identical columns in $H_E$. If these three identical columns are selected to construct $H_P$, each row weight of $H_P$ is either zero or three. Thus $\mathcal{C}$ cannot be 3 combinable.\\
(Only if part): For a parity-check matrix $H$ of a 3 combinable code $\mathcal{C} $with $n = 2M$, suppose that all the nonzero $m$-tuple column vectors do not appear exactly twice in $H$. Then $H_E$ should have at least three identical columns. If these three identical columns are selected to construct $H_P$, it contradicts the assumption that $\mathcal{C}$ is 3 combinable.\\
(If part): Clearly, $n=2M$ because the number of nonzero $m$-tuple column vectors is $M=2^m-1$. If we construct $H_P$ by choosing any three columns from $H_E$, there should be at least one distinct column from the others and therefore we can always find a row of weight one or two in $H_P$.
\end{proof}\vspace{6mm}

From Lemmas 4 and 5, the following theorem can be stated without proof.\\
\begin{theorem}
The combined-decodability of SPC codes is two.
\end{theorem}\vspace{6mm}

In order to find the combined-decodability of Hamming codes, the following lemmas are needed.\\
\begin{lemma}
If a vertical code $\mathcal{C}$ with an $m \times n$ parity-check matrix $H$ is 4 combinable, then $n \leq 3M$, where $M=2^m -1$. Moreover, when $n=3M$, $\mathcal{C}$ is $4$ combinable if and only if $H$ is constructed by using every nonzero $m$-tuple column vector exactly three times as its columns.
\end{lemma}\vspace{6mm}
\begin{proof}
If $n \geq 3M +1$, there should exist four identical columns in $H_E$. If these four identical columns are selected to construct $H_P$, each row weight of $H_P$ is either zero or four. Thus $\mathcal{C}$ cannot be 4 combinable.\\
(Only if part):
For a parity-check matrix $H$ of a 4 combinable code $\mathcal{C}$ with $n = 3M$ , suppose that all the nonzero $m$-tuple column vectors do not appear exactly three times in $H$. Then $H_E$ should have at least four identical columns. If those four identical columns are selected to construct $H_P$, there is no row of weight one or two in $H_P$. Thus it contradicts the assumption that $\mathcal{C}$ is 4 combinable.\\
(If part): Since each nonzero $m$-tuple column vector appears exactly three times in $H$, $n=3M$. If we choose any four columns from $H_E$, there should be at least one distinct column from the others. 

Suppose that there is neither a weight-1 row nor a weight-2 row in $H_P$ for $\epsilon =4$. Since both $H_E$ and $H_P$ are closed under the row-wise addition, $H_P$ can be of two distinct types, that is, the first type has $\begin{bmatrix} 1 &1 &1 &1 \end{bmatrix}$ as its nonzero rows and the second type has one 4-tuple of weight three, say $\begin{bmatrix} 1 &1 &1 &0 \end{bmatrix}$, as its nonzero rows such that
\begin{equation*}
	H_P = 
	\begin{bmatrix}
		0 & 0 & 0 & 0 \\[-3mm]
		\vdots & \vdots & \vdots & \vdots \\[-3mm]
		0 & 0 & 0 & 0 \\[-3mm]
		1 & 1 & 1 & 1 \\[-3mm]
		\vdots & \vdots & \vdots & \vdots \\[-3mm]
		1 & 1 & 1 & 1 \\
	\end{bmatrix} ~\textrm{or}~
	\begin{bmatrix}
		0 & 0 & 0 & 0 \\[-3mm]
		\vdots & \vdots & \vdots & \vdots \\[-3mm]
		0 & 0 & 0 & 0 \\[-3mm]
		1 & 1 & 1 & 0 \\[-3mm]
		\vdots & \vdots & \vdots & \vdots \\[-3mm]
		1 & 1 & 1 & 0 \\
	\end{bmatrix}.
\end{equation*}
However, note that the latter matrix cannot be obtained because there is no all-zero column in $H_P$. Therefore the former matrix is the only possible matrix for $H_P$ to have rows of weight zero or four.
 
Based on this result, we can see that there should be a row of weight one or two in $H_P$.
\end{proof}\vspace{6mm}
The following corollary is given without proof.\\
\begin{corollary}
If a vertical code $\mathcal{C}$ is 3 combinable, then $\mathcal{C}$ is also 4 combinable and thus $\mathcal{C}$ is 4 combined-decodable.
\end{corollary}\vspace{6mm}
\begin{lemma}
Suppose that vertical code $\mathcal{C}$ has minimum distance $d_{min}$ which is larger than or equal to 3. If $\mathcal{C}$ with an $m \times n$ parity-check matrix $H$ is 5 combinable, then $n \leq M$, where $M=2^m -1$. Moreover, when $n=M$, $\mathcal{C}$  is 5 combinable if and only if $H$ is constructed by using every nonzero $m$-tuple column vector only once as its columns.
\end{lemma}\vspace{6mm}
\begin{proof}
Since $d_{min}$ is larger than or equal to 3, $H$ should not contain identical columns. Therefore, $n \le M$ is trivial and the first part is proved. \\
(Only if part): For $n=M$ and $d_{min} \ge 3$, $H$ is uniquely constructed by using every nonzero $m$-tuple column vector only once as its columns.\\
(If part): Suppose that there is neither a weight-1 row nor a weight-2 row in $H_P$ for $\epsilon =5$. Since both $H_E$ and $H_P$ are closed under the row-wise addition, there are only two types of $H_P$ such as 
\begin{enumerate}
	\item $H_P$ should contain $\begin{bmatrix} 1 &1 &1 &1 &1 \end{bmatrix}$ as its only nonzero rows;
	\item $H_P$ should contain $\begin{bmatrix} 1 &1 &1 &1 &0 \end{bmatrix}$, $\begin{bmatrix} 1 &1 &0 &0 &1 \end{bmatrix}$, and $\begin{bmatrix} 0 &0 &1 &1 &1 \end{bmatrix}$ at least once as its nonzero rows.
\end{enumerate}
Note that the second type represents all the cases of three nonzero rows having 1, 2, and 2 zeros in distinct positions, respectively.
Then we have   
\begin{equation*}
	H_P = 
	\begin{bmatrix}
		0 & 0 & 0 & 0 & 0 \\[6mm]
		\vdots & \vdots & \vdots & \vdots & \vdots \\[6mm]
		0 & 0 & 0 & 0 & 0 \\[-3mm]
		1 & 1 & 1 & 1 & 1 \\[6mm]
		\vdots & \vdots & \vdots & \vdots & \vdots \\[6mm]
		1 & 1 & 1 & 1 & 1 \\
	\end{bmatrix} ~~\textrm{or}~~
	\begin{bmatrix}
		0 & 0 & 0 & 0 & 0 \\[-3mm]
		\vdots & \vdots & \vdots & \vdots & \vdots \\[-3mm]
		0 & 0 & 0 & 0 & 0 \\[-3mm]
		1 & 1 & 1 & 1 & 0 \\[-3mm]
		\vdots & \vdots & \vdots & \vdots & \vdots \\[-3mm]
		1 & 1 & 1 & 1 & 0 \\[-3mm]
		1 & 1 & 0 & 0 & 1 \\[-3mm]
		\vdots & \vdots & \vdots & \vdots & \vdots \\[-3mm]
		1 & 1 & 0 & 0 & 1 \\[-3mm]
		0 & 0 & 1 & 1 & 1 \\[-3mm]
		\vdots & \vdots & \vdots & \vdots & \vdots \\[-3mm]
		0 & 0 & 1 & 1 & 1 \\
	\end{bmatrix}.
\end{equation*}
Therefore, in order for $\mathcal{C}$ to be 5 combinable, $H$ should contain neither five identical columns nor two distinct pairs of columns.

If $H$ is constructed by using every nonzero $m$-tuple column vector only once as its columns, it is clear that $H$ has neither five identical columns nor two distinct pairs of columns and $\mathcal{C}$ is 5 combinable. Thus the proof is done.\\
%(Only if part): Suppose that a vertical code $\mathcal{C}$ with an $m \times (M+3)$ $H$ is 5 combinable. Then, from the above argument, $H$ should contain neither five identical columns nor two distinct pairs of columns. Therefore, the proof is clear. \\%For a parity-check matrix $H$ with $n = M+3$ of a 5 combinable code $\mathcal{C}$, if the other construction methods exist, there should exist 5 identical columns or two or more pairs of the same columns in $H$. Then, $\mathcal{C}$ should not be 5 combinable.\\
%(If part): Since one nonzero column of $H$ appears four times and the remaining $M-1$ nonzero columns appear only once, it is clear that $H_P$ cannot be any of the above two types. Therefore, $\mathcal{C}$ should be 5 combinable.
\end{proof}
\vspace{6mm}

For 5 combinable vertical code $\mathcal{C}$ without restriction on the minimum distance, it is not difficult to verify that the upper bound in Lemma 9 becomes $n \le 2M+2$. This bound can be proved by similar method used in the proof of Lemma 9.

\subsection{Combined-Decodability of Hamming Codes}
In this subsection, we will derive the combined-decodability of Hamming codes.

\begin{lemma}
Hamming codes as vertical codes are neither 6 combinable nor 7 combinable.
\end{lemma}\vspace{6mm}
\begin{proof}
For a $(7,4)$ Hamming code, $H_E$ is given in Subsection~\ref{sec:ExampleHamming}.
%\begin{equation*}
%	H_E = 
%	\begin{bmatrix}
%		1 & 0 & 1 & 1 & 1 & 0 & 0 \\[-3mm]
%		0 & 1 & 0 & 1 & 1 & 1 & 0 \\[-3mm]
%		0 & 0 & 1 & 0 & 1 & 1 & 1 \\[-3mm]
%		1 & 0 & 0 & 1 & 0 & 1 & 1 \\[-3mm]
%		1 & 1 & 0 & 0 & 1 & 0 & 1 \\[-3mm]
%		1 & 1 & 1 & 0 & 0 & 1 & 0 \\[-3mm]
%		0 & 1 & 1 & 1 & 0 & 0 & 1 \\
%	\end{bmatrix}.
%\end{equation*}
The first three rows in $H_E$ show all possible nonzero column vectors of length 3 and the rows in $H_E$ are all possible linear combinations of the first three rows except all-zero row. Since all rows in $H_P$ have weight four, a $(7,4)$ Hamming code is not 7 combinable.

For a $(2^m-1,2^m-m-1)$ Hamming code, the parity-check matrix consists of all nonzero $m$-tuple column vectors. Using only column permutation, we can reorder the parity-check matrix as
\begin{equation*}
	H = 
	\begin{bmatrix}
		0 & 0 & 0 & 0 & 0 & 0 & 0 & \cdots \\[-3mm]
		\vdots & \vdots & \vdots & \vdots & \vdots & \vdots & \vdots & \cdots \\[-3mm]
		0 & 0 & 0 & 0 & 0 & 0 & 0 & \cdots \\[-3mm]
		1 & 0 & 1 & 1 & 1 & 0 & 0 & \cdots \\[-3mm]
		0 & 1 & 0 & 1 & 1 & 1 & 0 & \cdots \\[-3mm]
		0 & 0 & 1 & 0 & 1 & 1 & 1 & \cdots \\[-3mm]
	\end{bmatrix}.
\end{equation*}
Note that the first seven column vectors have zeros in the first $m-3$ positions and they should be all different from each other. Since there is no all-zero column vector in $H$, there should appear all nonzero 3-tuple column vectors in the last three rows corresponding to the first seven columns. Then, we can construct $H_E$ by using all possible non-trivial linear combinations of rows in $H$, and $H_P$ by selecting the first seven columns from $H_E$.
It is clear that such $H_P$ is made up of all-zero rows and the rows of $H_E$ for a $(7,4)$ Hamming code. 

Since the weight of rows of $H_P$ is either 0 or 4, Hamming codes are not 7 combinable. It is also easy to check that Hamming codes are not 6 combinable because we can always find an $H_P$ for $e = 6$ which consists of weight-0, weight-3, and weight-4 rows by choosing any 6 columns from the above $H_P$ for $e = 7$.
\end{proof}\vspace{6mm}

%The following theorem is given without proof.\\
\begin{theorem}
The combined-decodability of Hamming codes as vertical codes is five. 
\end{theorem}\vspace{6mm}

\begin{proof}
From Lemmas 3, 4, 5, 7 and 9, Hamming codes are 5 combined-decodable. The theorem is directly proved by this result and Lemma 10.
\end{proof}\vspace{6mm}

\begin{table}[th!]
	\centering
	\caption{Distribution of $H_P$ for all possible horizontal codeword errors in the proposed decoding scheme with Hamming codes}
	\label{tab:distributionHamming}
	\begin{tabular}{|c|c|c|c|c|}\hline 
	\multirow{3}{*}{$(n,k)$}		& \multirow{3}{*}{$e$}	& \multicolumn{3}{c|}{Number of distinct $H_P$ for $e$}	\\ \cline{3-5}
								& 						& \multirow{2}{*}{Total $\binom{n}{e}$}& With rows of Hamming	& Without rows of Hamming	\\ 
								&						&						& weight one or two		& weight one or two			\\ \hline \hline
	\multirow{5}{*}{$(7,4)$}	& 6						& 7						& 0					& 7 					\\ \cline{2-5}
								& 5						& 21					& 21				& 0 					\\ \cline{2-5}
								& 4						& 35					& 35				& 0						\\ \cline{2-5}
								& 3						& 35					& 35				& 0						\\ \cline{2-5}
								& 2						& 21					& 21				& 0						\\ \hline
	\multirow{8}{*}{$(15,11)$}	& 10					& 3003					& 0					& 3003					\\ \cline{2-5}
								& 9						& 5005					& 396				& 4609					\\ \cline{2-5}
								& 8						& 6435					& 2772				& 3663					\\ \cline{2-5}
								& 7						& 6435					& 5313				& 1122					\\ \cline{2-5}
								& 6						& 5005					& 4900				& 105					\\ \cline{2-5}
								& 5						& 3003					& 3003				& 0						\\ \cline{2-5}
								& 4						& 1365					& 1365				& 0						\\ \cline{2-5}
								& 3						& 455					& 455				& 0						\\ \hline
	\multirow{8}{*}{$(31,26)$}	& 10					& 44352165				& 20827060			& 23525105				\\ \cline{2-5}
								& 9						& 20160075				& 15860120			& 4299955				\\ \cline{2-5}
								& 8						& 7888725				& 7681550			& 207175				\\ \cline{2-5}
								& 7						& 2629575				& 2597965			& 31610					\\ \cline{2-5}
								& 6						& 736281				& 735196			& 1085					\\ \cline{2-5}
								& 5						& 169911				& 169911			& 0						\\ \cline{2-5}
								& 4						& 31465					& 31465				& 0						\\ \cline{2-5}
								& 3						& 4495					& 4495				& 0						\\ \hline
	\multirow{5}{*}{$(63,57)$}	& 7						& 553270671				& 552631671			& 639000				\\ \cline{2-5}
								& 6						& 67945521				& 67935755			& 9766					\\ \cline{2-5}
								& 5						& 7028847				& 7028847			& 0						\\ \cline{2-5}
								& 4						& 595665				& 595665			& 0						\\ \cline{2-5}
								& 3						& 39711					& 39711				& 0						\\ \hline
	\end{tabular}
	\vspace{-8mm}
\end{table}
\begin{table}[th!]
	\centering
	\caption{Distribution of $H_P$ for all possible horizontal codeword errors in the proposed decoding scheme with DPC codes}
	\label{tab:distributionNonHamming}
	\begin{tabular}{|c|c|c|c|c|}\hline 
	\multirow{3}{*}{$(n,k)$}		& \multirow{3}{*}{$e$}	& \multicolumn{3}{c|}{Number of distinct $H_P$ for $e$}	\\ \cline{3-5}
								& 				& \multirow{2}{*}{Total $\binom{n}{e}$}& With rows of Hamming	& Without rows of Hamming \\ 
								&				&						& weight one or two		& weight one or two \\ \hline \hline
	\multirow{4}{*}{$(6,4)$}	& 5				& 6				& 0								& 6			\\ \cline{2-5}
								& 4				& 15			& 15							& 0				\\ \cline{2-5}
								& 3				& 20			& 20							& 0				\\ \cline{2-5}
								& 2				& 15			& 15							& 0				\\ \hline					
	\multirow{6}{*}{$(12,10)$}	& 7				& 792			& 0								& 792			\\ \cline{2-5}
								& 6				& 924			& 84							& 840			\\ \cline{2-5}
								& 5				& 792			& 360							& 432			\\ \cline{2-5}
								& 4				& 495			& 492							& 3			\\ \cline{2-5}
								& 3				& 220			& 208							& 12			\\ \cline{2-5}
								& 2				& 66			& 66							& 0				\\ \hline
	\end{tabular}
\end{table}
\begin{table}[h!]
	\centering
	\caption{Candidates for the vertical codes in the proposed decoding scheme}
	\label{tab:candidates}
	\begin{tabular}{|c|c|c|c|c|c|}\hline 
		\multirow{2}{*}{$(n,k)$}& \multirow{2}{*}{Code}		& Generator					& \multirow{2}{*}{$d_{\rm min}$}& $\tau$ 			& $\eta$	\\ 
								& 							& polynomial				& 								& correctability	& combined-decodability	\\ \hline \hline
		$(63,57)$				& \multirow{4}{*}{Hamming}	& $x^6+x+1$					& \multirow{4}{*}{3}			& \multirow{4}{*}{2}& \multirow{4}{*}{5}\\ \cline{1-1,3-3}
		$(31,26)$				& 							& $x^5+x^2+1$				& 								&					&					\\ \cline{1-1,3-3}
		$(15,11)$				& 							& $x^4+x+1$					& 								&					&					\\ \cline{1-1,3-3}
		$(7,4)$					& 							& $x^3+x+1$					& 								&					&					\\ \hline
		$(12,10)$				& \multirow{2}{*}{DPC}		& \multirow{2}{*}{$x^2+x+1$}& \multirow{3}{*}{2}			& \multirow{3}{*}{1}& 2					\\ \cline{1-1,6-6}
		$(6,4)$					& 							& 							& 								&					& 4					\\ \cline{1-3,6-6}
		$(k+1,k)$				& SPC						& $x+1$						& 								& 					& 2					\\ \hline
	\end{tabular}
	\vspace{-8mm}
\end{table}

Table~\ref{tab:distributionHamming} shows the distribution of $H_P$ for all possible unsuccessfully decoded horizontal codewords when Hamming codes are used as vertical codes. It clearly shows that combined-decodability of any Hamming code is five. Table~\ref{tab:distributionNonHamming} shows the distribution of $H_P$ for all possible unsuccessfully decoded horizontal codewords when double parity-check (DPC) codes are used as vertical codes. It shows that the combined-decodabilities of $(6,4)$ and $(12,10)$ DPC codes are four and two, respectively. To analyze the combined-decodability of the proposed decoding scheme, some simple linear codes such as SPC, DPC, and Hamming codes are considered as vertical codes, the parameters of which are listed in Table~\ref{tab:candidates}.

\section{Simulation Results}\label{sec:SimulationResults}
In this section, numerical analysis of the proposed decoding scheme is performed. The performance of LDPC codes with $R=1/2$ and $R=5/6$ in two different lengths $n=2304$ and $n=1152$ in IEEE~802.16e standard\cite{WMAN} are compared with that of the proposed decoding scheme. The maximum number of iterations for iterative belief propagation (BP) decoding is set to 50 for the entire simulations.

For the proposed decoding scheme, the $(24,23)$ SPC code is selected as the vertical code. To maintain the same overall code rate, the LDPC codes obtained by puncturing the above codes in IEEE~802.16e standard are used as the horizontal codes. By using the puncturing rate $1/24$, the overall code rates for the proposed decoding scheme are also set to $1/2$ and $5/6$.

\begin{figure}[ht!]
  	\centering
    \subfigure[$R=1/2$.]{
        \includegraphics[width=0.86\columnwidth]{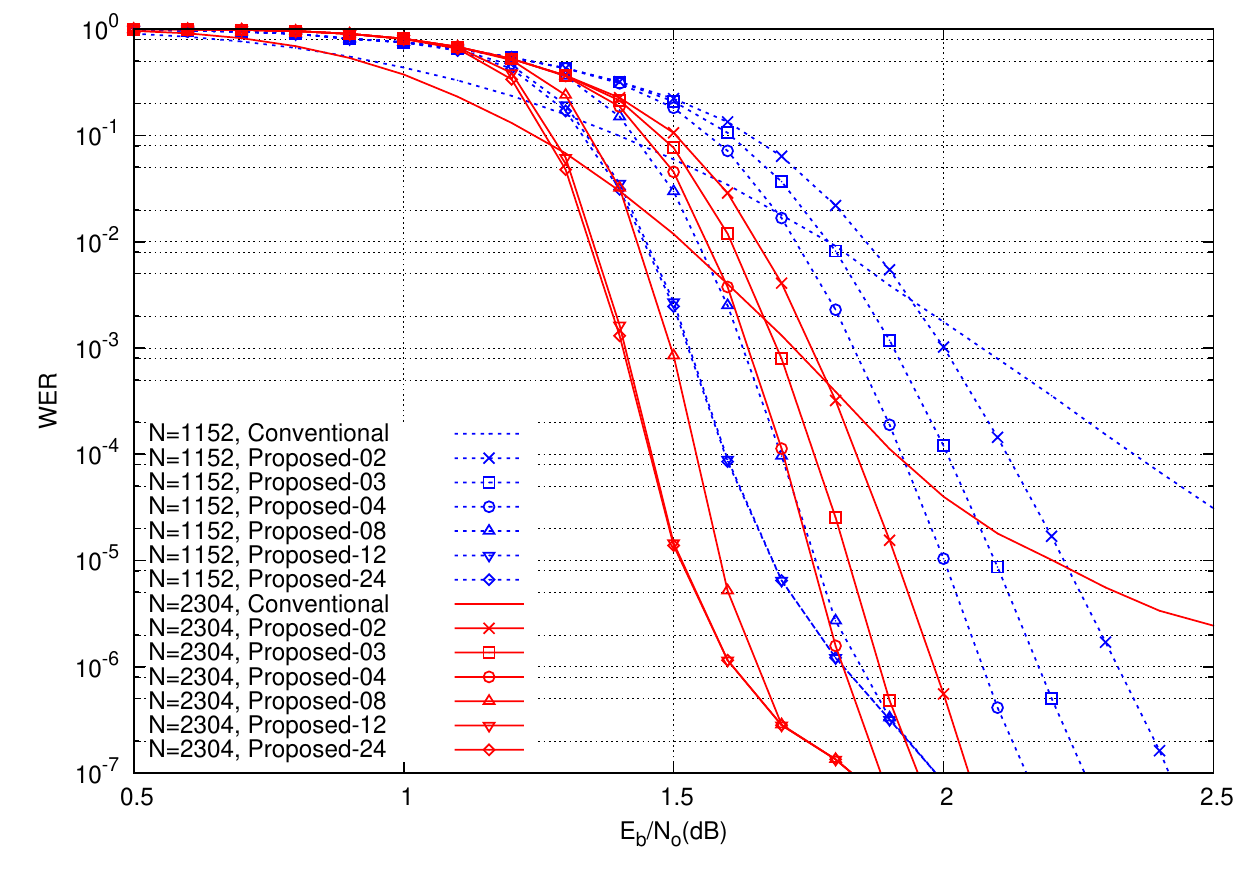}%
        \label{fig:Performance12}}\\
    \subfigure[$R=5/6$.]{
        \includegraphics[width=0.86\columnwidth]{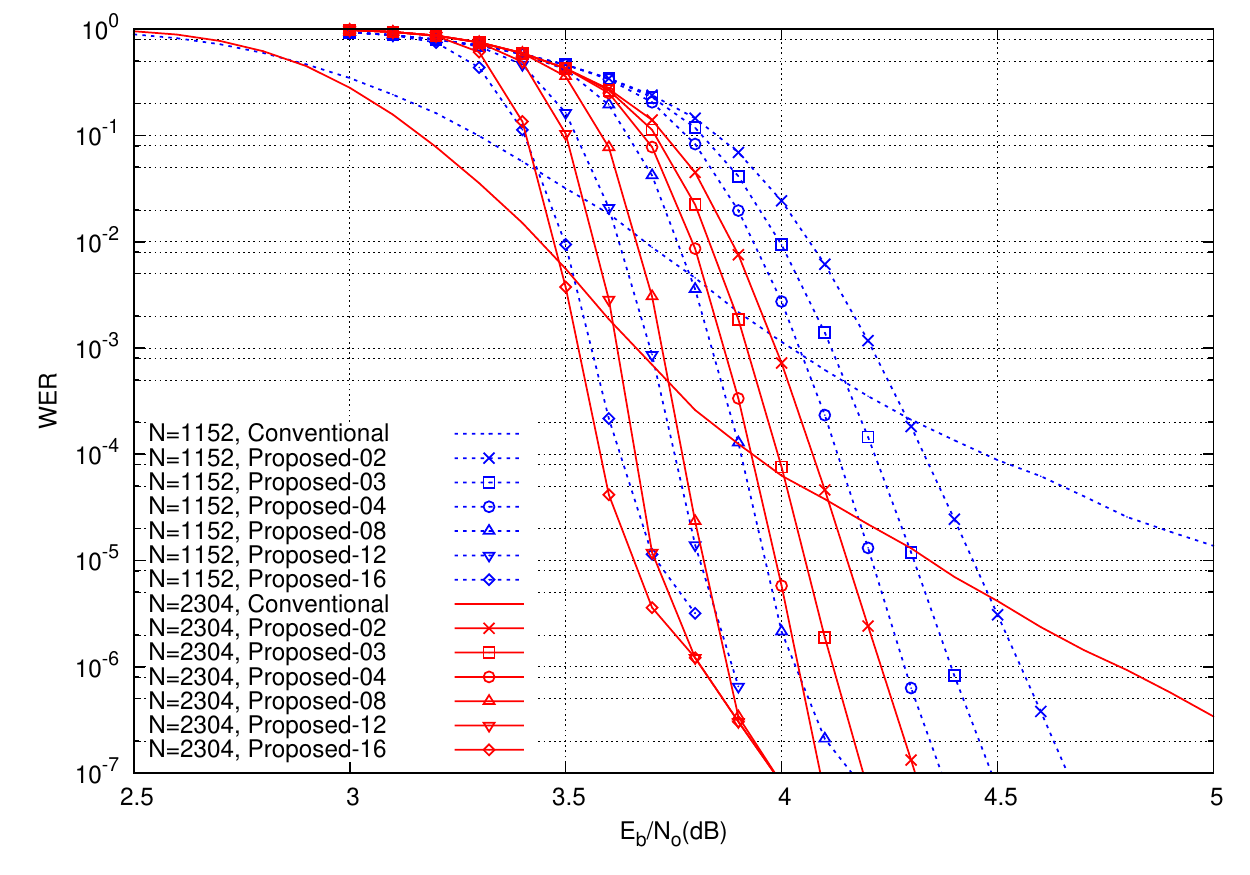}%
        \label{fig:Performance56}}
\caption{Performance comparison of the conventional decoding of LDPC codes in IEEE802.16e and the proposed decoding with $(24,23)$ SPC code as a vertical code.}
    \label{fig:Performance}
\end{figure}

Fig.~\ref{fig:Performance} shows that the proposed decoding scheme with $(24,23)$ SPC vertical code outperforms the conventional LDPC codes in IEEE802.16e standard for $R=1/2$ and $R=5/6$, especially in high SNR region, where `Proposed-$e$' means that up to $e$ horizontal codeword errors are combined-decoded. Error correction performance of each decoding scheme is shown in Fig.~\ref{fig:Performance} by word error rate (WER) of LDPC codewords for the conventional scheme and horizontal LDPC codewords for the proposed scheme. Even though the proposed decoding scheme does not correct all error patterns of $e$ or less horizontal codeword errors, Fig.~\ref{fig:Performance} shows remarkable performance improvement. The simulation results also show that both the waterfall and error-floor performance of the LDPC code can be improved by constructing a product code and performing the proposed decoding scheme.

Additional computational complexity for the proposed decoding scheme may be significant for low SNR region. However, for high SNR region, decoding failure of more than one horizontal codeword in a codeword matrix is rarely occurred. %And decoding for a codeword matrix with one unsuccessfully decoded horizontal codeword is simple in the proposed scheme. 
Thus, in spite of remarkable performance improvement, additional computational complexity per codeword becomes negligible for high SNR region. 

We showed the numerical results only for the case of SPC codes being used as vertical codes. Also, some of possible grouping methods for the erroneous horizontal codewords, i.e., grouping into one codeword and others, are considered for the case of $e_{\rm min}\ge 3$. If we use more powerful codes such as Hamming codes as vertical codes or consider more various grouping methods for the erroneous horizontal codewords, the performance of the proposed decoding scheme is believed to be improved but their decoding complexity increases.

%For the case of Hamming codes as vertical codes, the performance of the proposed decoding scheme is believed to be substantially improved compared to that of the conventional decoding scheme. But we cannot simulate this case due to the computational complexity and the variety of decoding cases.

%Fig.~\ref{fig:PerformanceHamming} shows the improved performance of the proposed decoding scheme with $(63,57)$ Hamming vertical code and horizontal codes punctured from LDPC codes in IEEE~802.16e standard, where `Case 2)' and `Case 3)' of the proposed decoding scheme are applied.

\section{Conclusion}\label{sec:Conclusions}
A new decoding scheme for LDPC codes using simple product code structure is proposed based on combining two independently received soft-decision data for the same codeword. The proposed decoding scheme can achieve very low error probability, which is very difficult to achieve only by using single channel coding scheme. Especially, the proposed decoding scheme can achieve better error-correcting capability in high SNR region. Also, the decoding capability of the proposed decoding scheme is analyzed using parity-check matrices of vertical codes and the {\it combined-decodability} is exactly derived when SPC and Hamming codes are used as vertical codes. 

It should be noted that the proposed decoding scheme can be applied to any other linear codes with soft-decision decoding than LDPC codes and will work well for the codes which rarely have undetected errors such as LDPC codes. %Modifying the proposed scheme to be efficient for the codes with many undetected errors is left as a further work.

%\section*{Acknowledgment}


\end{document}